\documentclass[10pt]{article}
\usepackage{mathrsfs}
\usepackage{graphicx}
\usepackage{bm}
\begin{document}

\title{Left-Right Polarization Asymmetry of \\the Weak Interaction Mass of Polarized Fermions in Flight}
\author{Zhi-Qiang Shi\thanks{E-mail address: zqshi@snnu.edu.cn}\\{\it \small Department of Physics, Shaanxi Normal University, Xi'an 710062, China P.R.}}

\date{}
\maketitle

\begin{abstract}
The left-right polarization-dependent asymmetry of the weak interaction mass is investigated. Based on the Standard Model, the calculation shows that the weak interaction mass of left-handed polarized fermions is always greater than that of right-handed ones in flight with the same speed in any inertial frame. The weak interaction mass asymmetry might be very important to the investigation of neutrino mass and would have an important significance for understanding the chiral attribute of weak interactions.\\



\end{abstract}

\section{Introduction}\label{1}\noindent

The Standard Model (SM) of particle physics is a chiral gauge theory and all the fermions are completely chiral, which necessarily results in the left-right polarization-dependent lifetime asymmetry. The lifetime asymmetry has been proposed based on theoretical analyzes \cite{1}. The relative ratio of decay probability for moving fermions is given by
\begin{equation}\label{eq:1}
    \frac{\Gamma_{_{Rh}}}{\Gamma_{_{Lh}}}=\frac{1-\beta}{1+\beta},
\end{equation}
where $\Gamma_{Rh}$ is the decay probability, caused by weak interaction, of the right-handed (RH) polarized fermions and $\Gamma_{Lh}$ is that of the left-handed (LH)
ones. The concrete calculation for decays of polarized muons has also been carried out within the framework of the SM \cite{2,3,4}. The result shows that LH
polarized muon lifetime $\tau_{_{Lh}}$ is different from RH polarized muon lifetime $\tau_{_{Rh}}$, i.e.
\begin{equation}\label{eq:2}
    \tau_{_{Lh}}=\frac{\tau}{1+\beta}\quad \hbox{and}\quad \tau_{_{Rh}}=\frac{\tau}{1-\beta},
\end{equation}
where $\tau$ is the average lifetime, $\tau=\gamma\tau_{_0}=\tau_{_0}/\sqrt{1-\beta^2}$, $\beta$ is the velocity of the muons measured in a natural system of units in
which $c=1$ and $\tau_{_0}$ is the lifetime in the muon rest frame. It is shown that the lifetime of RH polarized muons is always greater than that of LH ones in
flight. The lifetime asymmetry is conveniently described by the quantity
\begin{equation}\label{eq:3}
    A(L)_{_{LR}}=\frac{\tau_{_{Rh}}-\tau_{_{Lh}}}{\tau_{_{Rh}}+\tau_{_{Lh}}}=\beta.
\end{equation}

``The polarization asymmetry will play a central role in precise tests of the standard model.'' \cite{5} The lifetime asymmetry is the polarization asymmetry in charged weak current processes mediated by $W^\pm$ exchange. All existing experiments of measuring lifetime can not yet demonstrate the lifetime asymmetry to be right or wrong and the reasons have been analyzed in detail. In order to directly demonstrate the lifetime asymmetry, some possible experiments on the decays of polarized muons have been proposed \cite{6}. However, the polarization asymmetry in neutral weak current processes mediated by $Z$ exchange has been investigated widely. The SLD experiment measured the left-right polarization asymmetry in $e^-e^+$ collisions, in which the electrons are polarized while the positrons are not \cite{7,8,9,10,11,12}. The E158 experiment measured the left-right polarization asymmetry in polarized electron-electron M{\o}ller scattering \cite{13,14}. Their results show that the integrated cross section of LH polarized electrons is greater than that of RH ones . Therefore these experiments at SLAC have already indirectly demonstrated the lifetime asymmetry \cite{6}.

The concept of mass is one of the most fundamental notions in physics, comparable in importance only to those of space and time \cite{15}, and has a significant impact
on modern physics, from the realm of elementary particles to the cosmology of galaxies. The elementary fermions, leptons and quarks, may have associated finite
self-interaction energies which may be responsible for part or all of their masses. Although the value of the masses of the elementary fermions is correlated with the
strength of their dominant self-interaction, nondominant self-interactions also must play a role \cite{16}. Therefore, the physical mass $m$ of a particle may be
expressed by
\begin{equation}\label{eq:4}
    m=m(b)+\Delta m(s)+\Delta m(em)+\Delta m(w),
\end{equation}
where $m(b)$ is the so-called bare mass, $\Delta m(s)$, $\Delta m(em)$ and $\Delta m(w)$ are the strong, the electromagnetic and the weak interaction mass, respectively. The physical mass $m$ is a quantity which can be measured experimentally and also called the mechanical mass, the gravitational mass or the inertial mass.

From the view point of relativity, an event is described by four dimensional coordinates, three spatial coordinates and one time coordinate. Because the time is also a
generalized coordinate, its corresponding generalized momentum is the energy of a particle and the mass is shown to be a form of the energy, the lifetime asymmetry
would necessarily result in the mass asymmetry.

In this paper, we would like to investigate the left-right polarization-dependent asymmetry of the weak interaction mass. The outline of the paper is as follows. In
Sec.~\ref{2} and \ref{3}, we calculate the weak interaction mass caused by neutral and charged weak current, respectively. The result shows that the weak interaction
mass of LH polarized fermions is always greater than that of RH ones. Sec.~\ref{4} explores the impact of the weak interaction mass asymmetry on neutrino mass. In last Section, the results above are briefly summarized and its significance is discussed.

\section{The Weak Interaction Mass Caused by Neutral Weak Current}\label{2}\noindent

\begin{figure}[h]
\includegraphics{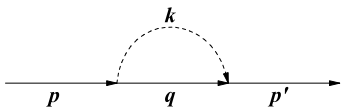}
\caption{\label{fig:mass-2}Weak self-energy graph of a fermion.}
\end{figure}

The weak interaction mass can be calculated by using the analogical method of calculating the electromagnetic mass. According to the SM, each elementary fermion may emit or absorb elementary gauge bosons connected with its elementary interactions either virtually or really, depending on energy. A free fermion with momentum $\bm p$ and
energy $E_p$, emitting and then absorbing a virtual neutral intermediate vector boson $Z$ with momentum $\bm k$ will acquire an energy which is known as weak
self-interaction energy. The intermediate state is comprised of a virtual boson $Z$ and a virtual fermion with momentum $\bm q$ as shown in Fig.~\ref{fig:mass-2}.

Considering weak coupling of Dirac fields, the total Hamiltonian for a weak interaction system is given by
\begin{equation}\label{eq:5}
    H=H_0+H_I,\quad H_0=H_D+H_B,
\end{equation}
where $H_D$ is the free Dirac field Hamiltonian, $H_B$ is the free boson field one and $H_I$ is the interaction one. The interaction Lagrangian density for neutral weak
current of elementary fermions is given by \cite{17}
\begin{equation}\label{eq:6}
    \mathscr{L}^Z_I(x)=Q\bar{g}\;\xi\;\bar{\psi}(x)\;\gamma_\mu\;\psi(x)\;Z_\mu(x)-\frac{1}{4}(-1)^{3Q}\;\bar{g}\;\bar{\psi}(x)\;\gamma_\mu\;(1+\gamma_5)\;\psi(x)\;Z_\mu(x),
\end{equation}
where $\bar{g}=\sqrt{g^2_{_1}+g^2_{_2}}$, $g_{_1}$ and $g_{_2}$ are the coupling constants corresponding to the groups U(1) and SU(2), respectively;
$\xi=\sin^2\theta_W=0.23$; $Q$ is charge number, $Q=0$ for neutrino, $Q=-1$ for electron, $Q=\frac{2}{3}$ for $u$-quark and $Q=-\frac{1}{3}$ for $d$-quark. The $\psi(x)$
is the plane wave solution of Dirac equation. $Z_\mu(x)$ is the neutral intermediate vector boson field. Therefore, the interaction Hamiltonian density is given by
\begin{equation}\label{eq:7}
    \mathscr{H}^Z_I(x)=-\mathscr{L}^Z_I(x).
\end{equation}
Obviously, the Hamiltonian density $\mathscr{H}^Z_I(x)$ comprises of two terms:
\begin{equation}\label{eq:8}
    \mathscr{H}_I^{(1)}(x)=-Q\bar{g}\;\xi\;\bar{\psi}(x)\;\gamma_\mu\;\psi(x)\;Z_\mu(x),
\end{equation}
and
\begin{equation}\label{eq:9}
    \mathscr{H}_I^{(2)}(x)=\frac{1}{4}(-1)^{3Q}\;\bar{g}\;\bar{\psi}(x)\;\gamma_\mu\;(1+\gamma_5)\;\psi(x)\;Z_\mu(x).
\end{equation}

In the SM, LH chirality state $\psi_{_{L}}(x)$ and RH chirality state $\psi_{_{R}}(x)$ are defined as, respectively,
\begin{equation}\label{eq:10}
  \psi_{_{L}}(x)=\!\frac{1}{2}(1+\gamma_5)\psi(x),\quad \psi_{_{R}}(x)=\!\frac{1}{2}(1-\gamma_5)\psi(x).
\end{equation}
We see from Eq.~(\ref{eq:8}) that $\mathscr{H}_I^{(1)}(x)$ is not related to the chirality of fermions and thereby can be absorbed into the expression of
unperturbed Hamiltonian $\mathscr{H}_0$ by means of the mass renormalization. However, it is not difficult to see from Eq.~(\ref{eq:9}) that $\mathscr{H}_I^{(2)}$
includes the chirality-state projection operator $(1+\gamma_5)$ which can only picks out LH chirality state in a spin state. The posterior discussion will show that it
is related to polarization and momentum of fermions. Therefore $\mathscr{H}_I^{(2)}$ can not be cancelled by means of the mass renormalization. We will concentrate on
its discussion. Based on Eq.~(\ref{eq:9}), the vertex function corresponding to $\mathscr{H}_I^{(2)}$ is given by
\begin{equation}\label{eq:11}
    i\frac{1}{4}(-1)^{3Q}\;\bar{g}\,(2\pi)^4\;\gamma_\mu\;(1+\gamma_5)\delta^{(4)}(p'-q-k).
\end{equation}

Now, let us calculate the self-energy process shown in Fig.~\ref{fig:mass-2}. Applying the Feynman rules, we easily find that the second order probability amplitude for
fermion weak self-energy transition from the initial state $|\bm p,s\rangle$ to the final state $|\bm p',s'\rangle$ is given by
\begin{eqnarray}
    \langle \bm p',s'|S^{(2)}|\bm p,s\rangle&=&-\frac{1}{16}\;\bar{g}^2\frac{1}{V}\frac{m}{E_p}\int d^4q\int d^4k\;
    \bar{u}_{s'}(p')\;\gamma_\mu(1+\gamma_5)\delta^{(4)}(p'-q-k)\nonumber\\
    &&\times\frac{-1}{\gamma\cdot q-im}\;\frac{-i\Big(g_{\mu\nu}-\frac{k_\mu k_\nu}{m^2_Z}\Big)}{k^2-m^2_Z}\;\gamma_\nu(1+\gamma_5)\delta^{(4)}(q+k-p)u_s(p)\nonumber\\
    &=&-i\frac{1}{8}(2\pi)^4\;\bar{g}^2\frac{1}{V}\frac{m}{E_p}\delta^{(4)}(p'-p)\;\bar{u}_{s'}(p')(1-\gamma_5)\Sigma(p)(1+\gamma_5)u_s(p),\label{eq:12}
\end{eqnarray}
where
\begin{equation}\label{eq:13}
    \Sigma(p)=\frac{-1}{(2\pi)^4}\int\frac{d^4k}{k^2-m^2_Z}\Bigg[\frac{\gamma\!\cdot\!(p-k)}{(p-k)^2+m^2}+\frac{1}{2m^2_Z}
    (\gamma\cdot k)\frac{\gamma\!\cdot\!(p-k)}{(p-k)^2+m^2}(\gamma\cdot k)\Bigg].
\end{equation}
The $u_s(p)$ is the plane wave solution of Dirac equation in momentum representation and known as spin state, in which $s$ is spin indices of spin states and $s=1,2$. In self-interaction process the state of external fermion remains unaltered, therefore Eq.~(\ref{eq:12}) can be rewritten as follows
\begin{equation}\label{eq:14}
    \langle \bm p,s|S^{(2)}|\bm p,s\rangle=-i\frac{1}{8}(2\pi)^4\;\bar{g}^2\frac{1}{V}\frac{m}{E_p}\;\delta^{(4)}(0)\bar{u}_{s}(p)(1-\gamma_5)\Sigma(p)(1+\gamma_5)u_s(p).
\end{equation}

However, we have pointed out emphatically that the polarization of fermions must be described by helicity states which are closely related to directly observable quantity experimentally \cite{2,3,4}. For LH and RH polarized fermions, therefore, substituting spin states in Eq.~(\ref{eq:14}) with helicity states, we obtain
\begin{equation}\label{eq:15}
    \langle \bm p,Lh|S^{(2)}|\bm p,Lh\rangle=-i\frac{1}{8}(2\pi)^4\;\bar{g}^2\frac{1}{V}\frac{m}{E_p}\;
    \delta^{(4)}(0)\bar{u}_{_{Lh}}(p)(1-\gamma_5)\Sigma(p)(1+\gamma_5)u_{_{Lh}}(p),
\end{equation}
\begin{equation}\label{eq:16}
    \langle \bm p,Rh|S^{(2)}|\bm p,Rh\rangle=-i\frac{1}{8}(2\pi)^4\;\bar{g}^2\frac{1}{V}\frac{m}{E_p}\;
    \delta^{(4)}(0)\bar{u}_{_{Rh}}(p)(1-\gamma_5)\Sigma(p)(1+\gamma_5)u_{_{Rh}}(p).
\end{equation}
A helicity state can be expanded as linear combination of chirality states:\cite{2,3,4}
\begin{equation}\label{eq:17}
    u_{_{Lh}}(p)=\sqrt{1+\beta}\;u_{_{L2}}(p^0)+\sqrt{1-\beta}\;u_{_{R2}}(p^0),
\end{equation}
\begin{equation}\label{eq:18}
    u_{_{Rh}}(p)=\sqrt{1-\beta}\;u_{_{L1}}(p^0)+\sqrt{1+\beta}\;u_{_{R1}}(p^0),
\end{equation}
where $u_{_{Lh}}(p)$ and $u_{_{Rh}}(p)$ are LH and RH helicity state, $u_{_{LS}}(p^0)$ and $u_{_{RS}}(p^0)$ are chirality state in the rest frame, respectively. $p^0$ is the four-momentum in the rest frame. Consequently we have
\begin{equation}\label{eq:19}
    (1+\gamma_5)u_{_{Lh}}(p)=\sqrt{1+\beta}(1+\gamma_5)u_2(p^0),
\end{equation}
\begin{equation}\label{eq:20}
    (1+\gamma_5)u_{_{Rh}}(p)=\sqrt{1-\beta}(1+\gamma_5)u_1(p^0),
\end{equation}
where $u_2(p^0)$ and $u_1(p^0)$ are spin states in the rest frame, respectively. Substituting Eqs~(\ref{eq:19}) and (\ref{eq:20}) into Eqs.~(\ref{eq:15}) and
(\ref{eq:16}), respectively, we obtain
\begin{equation}\label{eq:25}
    \langle \bm p,h|S^{(2)}|\bm p,h\rangle=-i\frac{1}{8}(2\pi)^4\;\bar{g}^2\frac{1}{V}\frac{m}{E_p}\delta^{(4)}(0)(1\pm\beta)\;
    \bar{u}_s(p^0)(1-\gamma_5)\Sigma(p)(1+\gamma_5)u_s(p^0),
\end{equation}
where plus sign refers to LH polarized fermions with $h=-1$ and $s=2$, while the minus sign to RH ones with $h=1$ and $s=1$.

To see the physical meaning of the $\langle \bm p,h|S^{(2)}|\bm p,h\rangle$, we note that for fermion weak self-energy transition process, generated by the interaction
Hamiltonian density
\begin{equation}\label{eq:26}
    \mathscr{H}_I^{\Delta m}(x)=-\Delta m(Z)\bar{\psi}(x)\psi(x),
\end{equation}
the transition matrix element is given by
\begin{equation}\label{eq:27}
    \langle \bm p,s|S^{({\Delta m})}|\bm p,s\rangle=-i(2\pi)^4\frac{1}{V}\frac{m}{E_p}\;\delta^{(4)}(0)\bar{u}_s(p)\Delta m(Z)_s\;u_s(p).
\end{equation}
For polarized fermions, according to Eqs.~(\ref{eq:17}) and (\ref{eq:18}), we obtain
\begin{eqnarray}
    \bar{u}_{_{Lh}}(p)u_{_{Lh}}(p)&=&\sqrt{1-\beta^2}\;\bar{u}_2(p^0)u_2(p^0),\label{eq:28}\\
    \bar{u}_{_{Rh}}(p)u_{_{Rh}}(p)&=&\sqrt{1-\beta^2}\;\bar{u}_1(p^0)u_1(p^0).\label{eq:29}
\end{eqnarray}
Thus Eq.~(\ref{eq:27}) can be rewritten as
\begin{equation}\label{eq:30}
    \langle \bm p,h|S^{({\Delta m})}|\bm p,h\rangle=-i(2\pi)^4\frac{1}{V}\frac{m}{E_p}\delta^{(4)}(0)\sqrt{1-\beta^2}\;\bar{u}_s(p^0)\Delta m(Z)_h\;u_s(p^0).
\end{equation}
Comparing the transition amplitude (\ref{eq:25}) with (\ref{eq:30}), we obtain
\begin{equation}\label{eq:31}
    \Delta m(Z)_h=\frac{1}{8}\;\bar{g}^2\frac{1\pm\beta}{\sqrt{1-\beta^2}}(1-\gamma_5)\Sigma(p)(1+\gamma_5).
\end{equation}
One can see that the self-interaction matrix element $\langle \bm p,h|S^{(2)}|\bm p,h\rangle$ amounts to a mass. The $\sqrt{1-\beta^2}$ in the denominator reflects the increase of mass with velocity in accordance with the principle of relativity.

Obviously, $\Sigma(p)$ is a divergent integral, as is easily seen by counting powers of $k$ in the numerator and denominator of the integrand in Eq.~(\ref{eq:13}). We
will be forced to study their relative values for avoiding the troubles they cause. The ratio of the weak interaction masses for RH and LH polarized fermions is given by
\begin{equation}\label{eq:32}
    \frac{\Delta m(Z)_{_{Rh}}}{\Delta m(Z)_{_{Lh}}}=\frac{1-\beta}{1+\beta}.
\end{equation}
The weak interaction mass asymmetry caused by neutral weak current is expressed by
\begin{equation}\label{eq:33}
    A(m)^Z_{_{LR}}=\frac{\Delta m(Z)_{_{Lh}}-\Delta m(Z)_{_{Rh}}}{\Delta m(Z)_{_{Lh}}+\Delta m(Z)_{_{Rh}}}=\beta.
\end{equation}

\section{The Weak Interaction Mass Caused by Charged Weak Current}\label{3}\noindent

The weak interaction mass may be caused by neutral or charged weak current. The former has been treated in above section. The latter will be treated in this section in which a virtual intermediate particle is charged boson $W_\mu(x)$.

The experiments and the theory have shown that neutral weak current is dominated by a coupling to both LH and RH chirality fermions and their weak coupling strengths are different, like Eq.~(\ref{eq:6}), while charged weak current by a coupling to LH chirality fermions and its interaction Lagrangian density is given by \cite{17}
\begin{equation}\label{eq:34}
    \mathscr{L}^W_I(x)=\frac{g_{_2}}{2\sqrt{2}}\;\bar{\psi}(x)\;\gamma_\mu\;(1+\gamma_5)\;\psi(x)\;W_\mu(x).
\end{equation}
Obviously, it is completely similar to Eq.~(\ref{eq:9}). Therefore, the ratio of the weak interaction masses for RH and LH polarized fermions is also similar to Eq.~(\ref{eq:32}), i.e.
\begin{equation}\label{eq:35}
    \frac{\Delta m(W)_{_{Rh}}}{\Delta m(W)_{_{Lh}}}=\frac{1-\beta}{1+\beta}.
\end{equation}
The weak interactions mass asymmetry caused by charged weak currents is expressed by
\begin{equation}\label{eq:36}
    A(m)^W_{_{LR}}=\frac{\Delta m(W)_{_{Lh}}-\Delta m(W)_{_{Rh}}}{\Delta m(W)_{_{Lh}}+\Delta m(W)_{_{Rh}}}=\beta.
\end{equation}

It can be see from Eqs.~(\ref{eq:32}) and (\ref{eq:35}) that the weak interaction mass of LH polarized fermions is always greater than that of RH ones in flight with the same velocity in any inertial frame. Synthesizing Eqs.~(\ref{eq:33}) and (\ref{eq:36}) the weak interaction mass asymmetry can uniformly be expressed by
\begin{equation}\label{eq:37}
    A(m)_{_{LR}}^w=A(m)_{_{LR}}^W=A(m)_{_{LR}}^Z=\beta.
\end{equation}
Comparing Eq.~(\ref{eq:37}) with Eq.~(\ref{eq:3}), we find out that the weak interaction mass asymmetry is similar to the lifetime asymmetry.

\begin{figure}[h]
\includegraphics{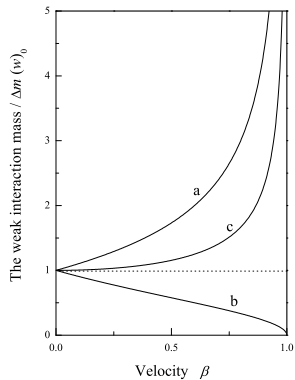}
\caption{\label{fig:mass-3} The weak interaction mass as a function of fermion velocity $\beta$. (a) The weak interaction mass $\Delta m(w)_{_{Lh}}$ of LH polarized
fermions. (b) The weak interaction mass $\Delta m(w)_{_{Rh}}$ of RH polarized fermions. (c) The weak interaction mass $\Delta m(w)$ of unpolarized fermions.}
\end{figure}

It should be noted that $\Delta m(w)$ is a polarization-dependent relativistic mass. From Eqs.~(\ref{eq:32}) and (\ref{eq:35}), the weak interaction mass can be rewritten as, respectively
\begin{equation}\label{eq:38}
     \Delta m(Z)_{_{Lh}}\sim \Delta m(W)_{_{Lh}}=\Delta m(w)_{_{Lh}}=(1+\beta)\;\Delta m(w)=(1+\beta)\;\gamma\;\Delta m(w)_0,
\end{equation}
\begin{equation}\label{eq:39}
     \Delta m(Z)_{_{Rh}}\sim \Delta m(W)_{_{Rh}}=\Delta m(w)_{_{Rh}}=(1-\beta)\;\Delta m(w)=(1-\beta)\;\gamma\;\Delta m(w)_0,
\end{equation}
where $\Delta m(w)_0$ is the weak interaction mass in the rest frame. As shown in Fig.~\ref{fig:mass-3}, when $\beta\neq 0$, the weak interaction mass $\Delta
m(w)_{_{Rh}}$ of RH polarized fermions is less than not only the weak interaction mass $\Delta m(w)_{_{Lh}}$ of LH ones but also $\Delta m(w)_{_0}$, and when the
velocity approaches to light speed, $\Delta m(w)_{_{Rh}}\rightarrow 0$.

\section{Neutrino Mass}\label{4}\noindent

There is a correlation between the mass of an elementary fermion and the relative strength of its dominant interaction. The mass of a hadron is mainly correlated with
strong interaction and mass of a charged lepton electromagnetic one. All elementary fermions take part in weak interaction. Because the strength of weak interaction
is much weaker than that of strong and electromagnetic interaction, the weak interaction mass is so small that it is unobservable and completely unimportant. However,
the detection of the weak interaction mass is not absolutely impossible. Long lifetime $K_L^0$ meson and short lifetime $K_S^0$ meson, for example, have different masses and $\Delta m=m_{K_L^0}-m_{K_S^0}=(3.483\pm 0.006)\times 10^{-6}$ eV \cite{18}. $K_L^0$ and $K_S^0$ are not charge conjugate state, having quite different decay modes
and lifetimes, so that, the mass difference $\Delta m$ --- but a very much smaller one --- should be attributed to difference in their weak coupling. In accompanying
with the progress of experimental technique and the improvement of the experimental accuracy, the detection of some extremely small effects would be becoming more and
more possible. The measurement of the neutrino oscillations is namely a very good example.

Besides gravitational interactions, neutrinos only take part in weak interactions and thereby its physical mass is regarded as partly bare mass and partly weak interaction mass without strong and electromagnetic mass. The interaction mass is proportional to interaction strength. The relative strength of electromagnetic and weak interactions are $10^{-2}$ and $10^{-5}$, respectively \cite{16}. The ratio of electromagnetic mass to electron mass (0.5 MeV) in the rest frame can be written
\begin{equation}\label{eq:40}
    \frac{\Delta m(em)_0}{m(e)_0}\sim \frac{10^{-2}}{0.5}=0.02,
\end{equation}
If electron neutrinos have 2 eV mass \cite{19}, then the ratio of weak interaction mass to electron neutrino mass in the rest frame can be written
\begin{equation}\label{eq:41}
    \frac{\Delta m(w)_0}{m(\nu_e)_0}\sim \frac{10^{-5}}{2\times 10^{-6}}=5.
\end{equation}
Obviously, the contribution of the weak interaction mass to electron neutrino mass would be greater than that of electromagnetic mass to electron mass.  For the massive neutrino, therefore, the effect of the weak interaction mass asymmetry would be very important.

In the SM, the neutrino has only the LH states, the antineutrino has only the RH states, and their masses must be zero. However, theoretically here is no compelling reason for massless neutrinos and the experiments of the neutrino oscillations have provided compelling evidences for nonzero neutrino masses and mixing \cite{20,21,22}. It will radically alter our understanding of the violation of parity conservation law and implies physics beyond the SM. In order to include massive neutrinos theoretically there have been numerous extensions of the SM containing neutrinos with Majorana or Dirac masses proposed in the literature. In these scenarios the most conspicuous version is the manifest left-right symmetric model, in which both LH and RH fields are treated in the same way---the LH neutrino currents are coupled with $SU(2)_L$ gauge boson, whereas the RH neutrino currents are coupled with $SU(2)_R$ gauge boson. It means that besides the standard $V-A$ current, there exist additional $V+A$ one mixed in the weak interactions. Also, there exist the interactions coupled with new neutral gauge boson $Z^{\prime}$.

For neutrinos, therefore, the interaction Hamiltonian density $\mathscr{H}_I^{(2)}(x)$, Eq.~(\ref{eq:9}), should be written as \cite{23}
\begin{eqnarray}\label{eq:42}
    \mathscr{H}_I^{(2)}(x)&\sim &J_\mu(x) Z_\mu(x) + J^{\prime}_\mu(x) Z^{\prime}_\mu(x)\nonumber\\
    &\sim &a_1\;\bar{\psi}(x)\;\gamma_\mu\;(1+\gamma_5)\;\psi(x)\;Z_\mu(x)+\;a_2\;\bar{\psi}(x)\;\gamma_\mu\;(1-\gamma_5)\;\psi(x)\;Z_\mu(x)\nonumber\\
    & + & a_1^{\prime}\;\bar{\psi}(x)\;\gamma_\mu\;(1+\gamma_5)\;\psi(x)\;Z^{\prime}_\mu(x)+\;a_2^{\prime}\;\bar{\psi}(x)\;\gamma_\mu\;(1-\gamma_5)\;\psi(x)\;Z^{\prime}_\mu(x),
\end{eqnarray}
where $a_1, a_2, a_1^{\prime}$ and $a_2^{\prime}$ are coupling coefficients. Applying the analogous method above to solve the weak interaction mass,  Eq.~(\ref{eq:37}) can be modified as
\begin{equation}\label{eq:43}
    A(m)_{_{LR}}^w=\alpha\beta,
\end{equation}
where
\begin{equation}\label{eq:44}
    \alpha=\frac{C^2_1-C^2_2}{C^2_1+C^2_2},
\end{equation}
$C_1^2=a_1^2+a_1^{\prime 2}$ and $C_2^2=a_2^2+a_2^{\prime 2}$. If $C_1=C_2$, then $\alpha=0$, there is no the weak interaction mass asymmetry. If $C_2=0$, then $\alpha=1$, there is no $V+A$ interaction, i.e., the weak interaction mass asymmetry reaches its maximum.

According to Eqs.~(\ref{eq:4}), (\ref{eq:38}) and (\ref{eq:39}), the motion mass of a neutrino in flight can be written as
\begin{equation}\label{eq:45}
    m(\nu)_{_{Lh}}=m(b)+\Delta m(w)_{_{Lh}}=\frac{m(b)_0}{\sqrt{1-\beta^2}}+(C^2_1+C^2_2)(1+\alpha\beta)\frac{\Delta m(w)_0}{\sqrt{1-\beta^2}},
\end{equation}
\begin{equation}\label{eq:46}
    m(\nu)_{_{Rh}}=m(b)+\Delta m(w)_{_{Rh}}=\frac{m(b)_0}{\sqrt{1-\beta^2}}+(C^2_1+C^2_2)(1-\alpha\beta)\frac{\Delta m(w)_0}{\sqrt{1-\beta^2}},
\end{equation}
where $m(b)_0$ is the bare mass of a neutrino in the rest frame. One sees that the motion mass of LH polarized neutrinos is always greater than that of RH ones. The asymmetry of the motion mass of a neutrino can be expressed by
\begin{equation}\label{eq:47}
    A(m(\nu))_{LR}=\frac{m(\nu)_{_{Lh}}-m(\nu)_{_{Rh}}}{m(\nu)_{_{Lh}}+m(\nu)_{_{Rh}}}=\alpha\beta\frac{\Delta m(w)_0}{m(\nu)_0},
\end{equation}
where $m(\nu)_0$ is the physical mass of a neutrino in the rest frame and $m(\nu)_0=m(b)_0+\Delta m(w)_0$. The mass asymmetry increases with the increase of neutrino's velocity and the ratio of $\Delta m(w)_0$ to $m(\nu)_0$.

\section{Summary and discussion}\label{5}\noindent

Within the framework of the SM in particle physics, we have proved the weak interaction mass of LH polarized fermions is always greater than that of RH ones in flight with the same speed in any inertial frame. The SLD and E158 experiment have indirectly demonstrated not only the lifetime asymmetry but also the weak interaction mass asymmetry. Therefore, we point out emphatically here that the idea of the weak interaction mass asymmetry has already been linked to some experimental discoveries, and it is not merely a guess without justification.

Although the weak interaction mass asymmetry is generally negligible, it might have a great impact on neutrino mass because its dominant self-interaction is weak. As see from Eqs.~(\ref{eq:45}), (\ref{eq:46}) and (\ref{eq:47}), the motion mass of LH polarized neutrino is always greater than that of RH one. Though neutrino has non-zero mass, its mass is so small that its velocity is very high and could even approach the speed of light. As shown in Fig.~\ref{fig:mass-3}, therefore, the weak interaction mass could hardly contribute to the mass of RH polarized neutrinos when $\beta\rightarrow 1$. This means that the motion mass of RH polarized neutrino would be extremely small relative to that of LH one. It might be the reason why RH neutrinos can not be found experimentally.

There is strong evidence for the existence of a substantial amount of dark matter which is a mixture of cold dark matter (CDM) containing 70 \% and hot dark matter (HDM)
30 \% \cite{24,25}. The foremost HDM candidate is neutrino. Because the fraction of LH neutrino contributing to the universe mass should be much greater than that of RH one, even in the presence of RH neutrinos, the HDM should still be a kind of ``polarized" or ``handedness" dark matter. It might have important consequences for the research of dark matter and the evolution of the universe.

The weak interaction mass asymmetry is inevitable outcome of the chiral attribute of weak interactions. The chiral attribute is experimentally exhibited the parity nonconservation in space, the lifetime asymmetry in time and the weak interaction mass asymmetry as an effect of space-time. The three asymmetrical phenomena have described the attribute from all aspects of the space and the time. Although the weak interaction mass asymmetry is quite small, it enabled us to more profoundly cognise and understand the dynamical structure of the weak interactions, the essence of the parity nonconservation as well as the characteristic of space-time structure.

\section*{acknowledgments}

I would like to thank Professor Guang-Jiong Ni for his many informative discussions and invaluable help.

\end{document}